\documentclass[twocolumn]{aastex631}
\hypersetup{linkcolor=blue,citecolor=blue,filecolor=blue,urlcolor=blue}

\usepackage{svg}
\usepackage{rotating}
\usepackage{appendix}
\usepackage{graphicx}
\usepackage{epstopdf}
\usepackage{natbib}
\usepackage{xspace}
\usepackage{color}
\usepackage{amsmath}

\newcommand{\mjyb}   {~mJy~beam$^{-1}$\xspace}
\newcommand{\kms}     {~km~s$^{-1}$\xspace}

\newcommand{\msun}    {~$M_{\sun}$\xspace}

\newcommand{\co}      {CO\,(3-2)\xspace}

\begin{document}

\title{Possible Explosive Dispersal Outflow in IRAS\,16076-5134 revealed with ALMA}

\author[0000-0003-2630-3774]{Estrella Guzmán Ccolque}
\affiliation{Instituto Argentino de Radioastronomía (CCT- La Plata, CONICET,CICPBA) \\
C.C. No. 5,1894, Villa Elisa, Buenos Aires, Argentina}
\email{estreguzman@gmail.com}

\author[0000-0001-5811-0454]{Manuel Fernández-López}
\affiliation{Instituto Argentino de Radioastronomía (CCT- La Plata, CONICET,CICPBA), C.C. No. 5,1894, Villa Elisa, Buenos Aires, Argentina}

\author[0000-0003-2343-7937]{Luis A. Zapata}
\affiliation{Instituto de Radioastronomía y Astrofísica. Universidad Nacional Autónoma de México, P.O. Box 3-72,58090, Morelia, Michoacán,México}

\author[0000-0003-0295-6586]{Tapas Baug}
\affiliation{S. N. Bose National Centre for Basic Sciences, Block-JD, Sector-III, Salt Lake, Kolkata 700106, India}

\begin{abstract}
We present 0.9\,mm continuum and \co line emission observations retrieved from the Atacama Large Millimeter/submillimeter Array (ALMA) archive toward the high-mass star formation region IRAS\,16076-5134. We identify fourteen dense cores with masses between 0.3 to 22\msun. We find an ensemble of filament-like \co ejections from $-$62 to +83\kms that appear to arise radially from a common central position, close to the dense core MM8. The ensemble of filaments, has a quasi-isotropic distribution in the plane of the sky. The radial velocity of several filaments follow a linear velocity gradient, incresing from a common origin. Considering the whole ensemble of filaments, we estimate its total mass to be 138 and 216\msun from its CO emission, for 70\,K and 140\,K respectively. Also, assuming a constant velocity expansion of the filaments (of 83\kms) we estimate the dynamical age of the outflowing material (3500 years), its momentum ($\sim10^{4}$\msun\kms) and its kinetic energy ($\sim10^{48-49}$\,erg). The morphology and kinematics presented by the filaments suggest the presence of a dispersal outflow with explosive characteristics in IRAS\,16076-5134. In addition, we make a raw estimate of the lower limit of the frequency rate of the explosive dispersal outflows in the Galaxy (one every 110 years) considering constant star formation rate and efficiency with respect to the galactocentric radius of the Galaxy. This may imply a comparable rate of dispersal outflows and supernovae (approximately one every 50 years), which may be important for the energy budget of the Interstellar Medium and the link between dispersal outflows and high-mass star formation.
\end{abstract}


\section{Introduction} \label{sec:intro}
IRAS\,16076-5134 or AGAL\,331.28-00.19 (hereafter I16076) is an ultracompact HII (UCHII) region located at 5.0$\pm$0.7\,kpc \citep[][ kinematic distance calculated using the Kinematic Distance Calculation Tool of \cite{2018_Wenger}]{2020_Baug}. This measure is similar to the value obtained previously by \cite{2004_Faundez} (5.2\,kpc), derived using a flat rotation model of the Milky Way and the CS(2-1) line velocity reported by \citep{1996_Bronfman}. I16076 has FIR color typical of UCHII regions \citep{1989_Wood}. \cite{2004_Faundez} detected its dust emission at 1.2\,mm, measuring its size ($38\arcsec$), dust temperature (33\,K) and mm-IR bolometric luminosity (2.0$\times10^{5}\,L_{\odot}$). Such large mm-IR bolometric luminosity in a cold dust region usually implies its link with a region forming one or more stars of spectral types B and O. A dense clump (radius=1.57\,pc) in I16076 was also observed in the ATLASGAL survey at 870\,$\mu$m for which \cite{2018_Urquhart} presented values of H$_{2}$ column density (23.3$\times10^{15}$\,cm$^{-2}$), mass (45.6\msun) and dust temperature (30.1\,K). In addition, they used a combination of archival observations from molecular line surveys reported in the literature (such as CO, NH$_3$, CS, etc) to determine the radial velocity with regard to the local standard of rest ($v_{lsr}$): -87.7\kms. I16076 shows strong blueshifted emission profile of HCN(4-3), known to be a tracer of infalling gas and therefore an indicator of accretion processes \citep{2016_Liu}. More recently, I16076 was observed with ALMA at 3\,mm in order to explore the molecular outflows in this region \cite{2020_Liu}. \cite{2020_Baug} detected several outflow lobes in \co toward IRAS16076. Most of them are associated with the millimeter sources identified by \cite{2021_Baug}. In particular, from the dense core identified as I16076\_O1 seem to arise 19 outflow lobes (9 redshifted and 10 blueshifted) distributed almost isotropically. This morphology is reminiscent of the dispersal outflows with explosive features detected in Orion BN/KL, DR21 and G5.89-0.39 \citep{2015_Bally,2017_Bally, 2017_Zapata,2019_Zapata,2020_Zapata}. Hereafter we refer to this type of outflows as \lq explosive dispersal outflows\rq or just \lq dispersal outflows\rq.

Since their discovery, explosive dispersal outflows have become a new subclass of molecular outflows observed in the star-forming regions of massive stars \citep[e.g.,][]{2009_Zapata}. At present, there seem to be two types of molecular outflows. The first type is the classical outflows which are typically bipolar, produced mostly by low-mass stars in their formation process \citep{2007_Arce}. Theoretical work suggests that this outflows are intimately related to the accretion process and play an important role in dissipating excess angular momentum in the infalling material and
allow the star to build up its mass through accretion \citep[e.g.,][]{1987_Shu,1991_Shu,2000_Shu,2000_Konigl}. On the other side, the dispersal outflows with explosive characteristics, whose nature is still unknown \citep{2005_Bally, 2009_Zapata,2011_Bally,2017_Bally}. They have been associated to an energy release produced by the disintegration or close encounters of young stellar systems \citep{2017_Bally,2021_Rivera-Ortiz}  or the merger of several massive protostars \citep{2005_Bally}. These outflows appear to be impulsive and created by a single brief energetic event with energy injections of about $10^{47-49}$\,erg \citep{2005_Bally}. They comprise dozens of filament-shaped ejections isotropically distributed on the sky. Each filament follows a well-defined Hubble velocity law, that is the radial velocities linearly increase with projected distances \citep{2009_Zapata,2013_Zapata,2017_Bally,2019_Zapata}. Dispersal outflows are usually detected by their CO emission \citep{2012_Peng,2017_Zapata,2019_Zapata,2020_Zapata} but they have also been detected in H$_{2}$ \citep{2011_Bally}, and [FeII] condensations have been observed at the tips of the filaments in the case of Orion BN/KL \citep{2015_Bally}. 

\cite{2017_Zapata} enumerated four clear morphological and kinematical features of explosive dispersal outflows that distinguishes them from classical bipolar molecular outflows:
\begin{itemize}
    \item The explosive dispersal outflows consist of narrow straight filament-like ejections with different orientations and an almost isotropic distribution.
    \item The filaments point back to common position where the explosive event may have occurred.
    \item The radial velocity of each filament increases linearly with distance projected on the sky from the origin, it seems to follow a Hubble's law. 
    \item The redshifted and blueshifted filaments seem to overlap in the plane of the sky.
\end{itemize}
Moreover:
\begin{itemize}
    \item These outflows are associated with regions of high-mass star formation.
    \item Their energies are in the range of $10^{47}-10^{49}$\,erg. 
\end{itemize} 

Using observations from the ALMA archive, and following the features of explosive dispersal outflows described in \citep{2017_Zapata} we propose the possible presence of this type of event in I16076. We reanalyze the dust and line kinematics of the ALMA data under this hypothesis, and present the new implications of the discovery. The paper is organized as follows. In Section \ref{sec:observaciones} we present the ALMA observations. The identification and derivation of the physical parameters of the 0.9\,mm continuum sources, as well as of the \co filamentary ejections are presented in Section \ref{sec:results}. A brief discussion on whether I16076 is an explosive dispersal outflow, and the rate of such events in the Galaxy can be found in Section \ref{sec:discussion}. The conclusions of this study are presented in Section \ref{sec:conclusion}.

\section{ALMA observations} \label{sec:observaciones}
The Band\,7 observations were carried out during ALMA (Atacama Large Millimeter/submillimeter Array) Cycle\,5, as part of the program 2017.1.00545.S (P.I: Tie, Liu). The array was in C43-2 configuration, with 48 12-m antennas available at the time, and a baseline ranging 15-314\,m. The I16076 data presented in this article correspond to a portion of a larger program, comprising 10 more sources. I16076 was observed in three sessions (Table \ref{tab:observations}) during May 2018, totaling approximately 21 minutes on source under good weather conditions, with precipitable water vapor (PWV) around 0.4\,mm. The observations were made in mosaic mode with 17 pointings covered during each individual session. The coordinates of the mosaic center were $(\alpha,\delta)_{J2000.0}=16^{h}11^{m}27.1^{s},-51^{\circ}41^{'}55.1^{''}$.

The correlator was configured to use four spectral windows centered at 343.302, 345.183, 354.505 and 356.840\,GHz. The former two were used to image the continuum and $^{12}$CO(3-2) emission, respectively; they had a 0.976\,MHz spectral resolution and a total bandwidth of 1.875\,GHz. The remaining spectral windows aimed at the HCN(4-3) and HCO(4-3) lines; they had a 0.244\,MHz (0.85\kms at the observing frequency) spectral resolution and a total bandwidth of 0.469\,GHz. In this work, we focus on the 0.9\,mm continuum and the \co line emission. J1650-5044 was used as the phase calibrator, J1427-4206 and J1517-2422 as the bandpass and flux calibrator. The estimation for the absolute flux calibration these frequency in ALMA is $\sim10\%$\footnote{See Section 10.4.7 in the ALMA Cycle 8 Technical Handbook}.

Visibility data were reduced using CASA\,5.1.1 (Common Astronomy Software Applications) package. First, the data were calibrated by running the calibration script provided by the ALMA staff. The images were constructed using \textit{tclean} task of the CASA version 6.1.2. The cleaning process was performed interactively applying the Hogbom minor cycle algorithm and Briggs weighting, setting the robust parameter to 0.5, which ensures a good compromise between the sensitivity and the angular resolution of the images. The rms of the continuum image is 1.3\mjyb and the angular resolution is $0\farcs8\times0\farcs7$ (PA$=74\degr.6$; where PA is the position angle \footnote{PA is the measured angle of the beam's major axis with respect to the north celestial pole, turning positive in the right ascension direction (i.e., toward East).}). For the CO cube the rms is 5.6\mjyb per channel and the angular resolution is $0\farcs9\times0\farcs8$ (PA=$80\degr.6$).

\begin{deluxetable*} {cccc}
\tabletypesize{\footnotesize}
\tablecolumns{4}
\tablewidth{0pt}
\tablenum{1}
\tablecaption{Summary of Band 7 ALMA Observations of IRAS16076-5134 (Project 2017.1.00545.S) \label{tab:observations} \\}
\tablehead{
\colhead{Execution Blocks IDs} & \colhead{uid://A002/Xcda49e/Xb713} & \colhead{uid://A002/Xcda49e/Xc0a3} & \colhead{uid://A002/Xcdb7b8/Xd8c8}
}
\startdata
Observation date & 2018-05-21 & 2018-05-21 & 2018-05-23 \\
Number of antennas & 48 & 48 & 45 \\
Time on Source (sec) & 428.2 & 428.3 & 428.2 \\
Mean PWV(mm) & 0.4 & 0.5 & 0.3 \\
Phase Calibrator & J1650-5044 & J1650-5044 & J1650-5044 \\
Bandpass Calibrator & J1427-4206 & J1427-4206 & J1517-2422 \\
Flux Calibrator & J1427-4206 & J1427-4206 & J1517-2422
\enddata
\end{deluxetable*}

%\begin{deluxetable*} {cccc}
%\tablenum{1}
%\tablecaption{Summary of Band 7 ALMA Observations of IRAS16076-5134 (Project 2017.1.00545.S) %\label{tab:observations}}
%\tablewidth{0pt}
%\tablehead{
%\colhead{Execution Blocks IDs} 
%\nocolhead{} & \colhead{uid://A002/Xcda49e/Xb713} & \colhead{uid://A002/Xcda49e/Xc0a3} & %\colhead{uid://A002/Xcdb7b8/Xd8c8}
%}
%\startdata 
%Observation date & 2018-05-21 & 2018-05-21 & 2018-05-23\\
%Number of antennas & 48 & 48 & 45\\
%Time on Source (sec) & 428.2 & 428.3 & 428.2\\
%Mean PWV(mm) & 0.4 & 0.5 & 0.3\\
%Phase Calibrator & J1650-5044 & J1650-5044 & J1650-5044\\
%Bandpass Calibrator & J1427-4206 & J1427-4206 & J1517-2422\\
%Flux Calibrator & J1427-4206 & J1427-4206 & J1517-2422\\
%\enddata
%\label{tabla_observaciones}
%\end{deluxetable*}

\section{RESULTS} \label{sec:results}
\subsection{Submillimeter Continuum Sources}

\begin{figure*}
\centering
\includegraphics[scale=0.45]{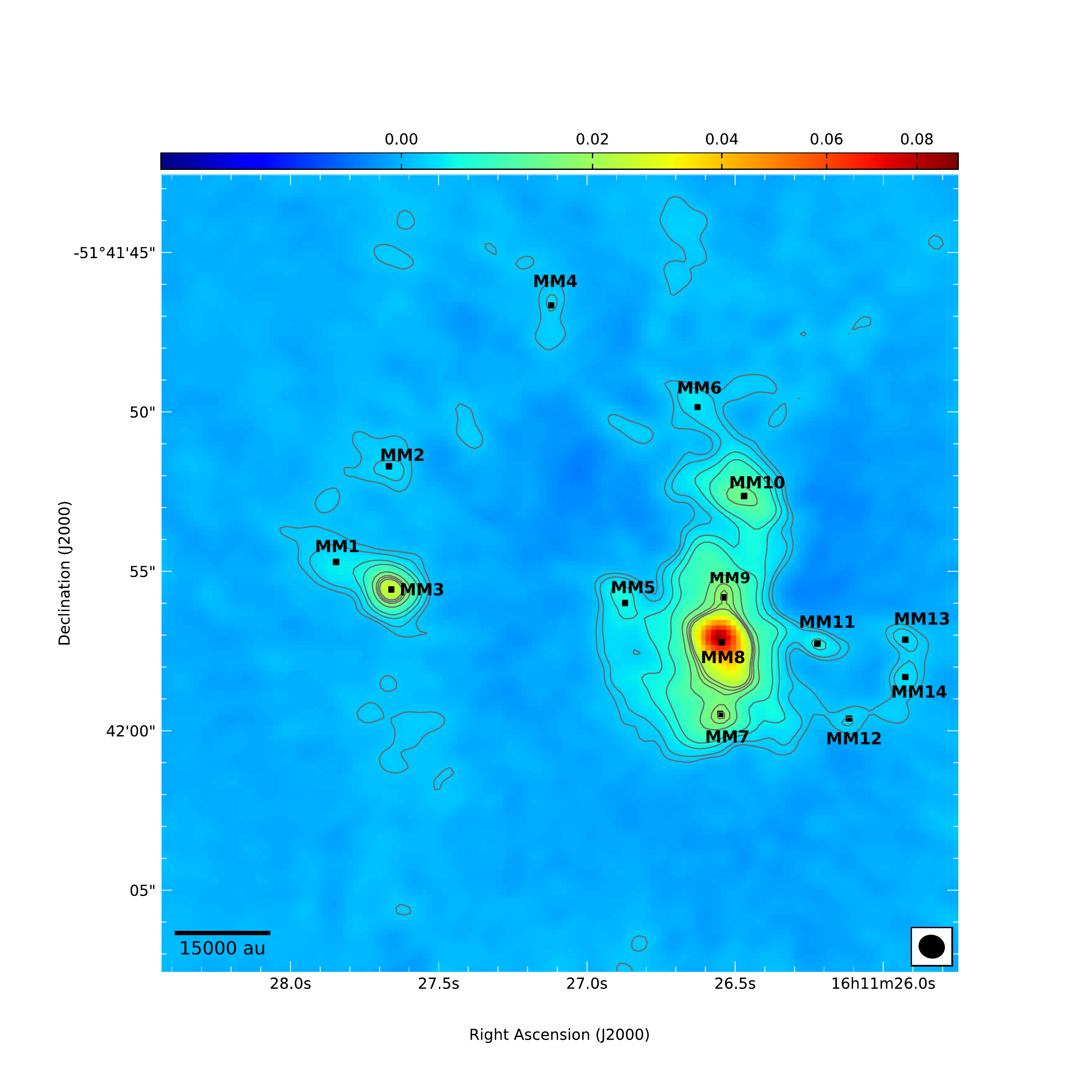}
\caption{0.9\,mm continuum emission toward IRAS\,16076-5143 in colour-scale and grey contours (at 3,5,10,15,30,40,45 times the rms noise level of 4\mjyb). The synthesized beam, in the bottom right corner, is $0\farcs8\times0\farcs7$, PA$=74\degr.6$. The submillimeter continuum sources identified here are marked by black dots.}
\label{fig:cores}
\end{figure*}

Figure \ref{fig:cores} shows the positions of fourteen 0.9\,mm continuum sources detected with ALMA in I16076. These sources are labeled as MM sources sorted by their Right Ascension. We consider a flux density threshold of 5$\sigma\sim$\,7\mjyb as a criterion to visually identify continuum peaks, which we associate with continuum sources (here $\sigma$ is estimated analyzing the empty background emission). The continuum map shows the sources are mainly distributed in two small clusters. One small cluster around MM3, and a larger one around MM8. Table \ref{tab:continuum} presents some parameters to characterize these continuum sources, including central coordinates, deconvolved sizes, peak intensities, flux densities and masses. These parameters, except for the mass, and their uncertainties were estimated fitting 2D-Gaussians by using the CASA task \texttt{imfit}. Since the sizes of the millimeter sources are of the order of $\sim$\,0.1\,pc, we assume that they correspond to dense cores\footnote{We use the nomenclature and definition in \cite{2009_Zhang}, \cite{2018_Motte}, \cite{2017_Bally} and \cite{2014_Hull}, who refer as clouds to 10-100\,pc structures, molecular clumps to 1\,pc structures where populations of both massive and lower mass stars are formed, dense cores to 0.01-0.1\,pc structures forming one or a group of stars; and protostellar envelopes to 1000\,AU ($\sim$\,0.005\,pc) structures comprising the densest parts of the dense core, inside which one or several protostars are formed.}. MM1, MM3, MM5, MM7, MM8 and MM9 were previously identified by \cite{2021_Baug}, who also considered these sources as dense cores.

We calculate the masses of the dense cores using: 
%$$M_{core} = \frac{F_{\nu} d^{2} R }{B_{\nu}(T_{d})\,\kappa_{\nu}}$$
\begin{equation} \label{mass_core}
M_{core} = \frac{F_{\nu} d^{2} R }{B_{\nu}(T_{d})\,\kappa_{\nu}}
\end{equation}
where $F_{\nu}$ is the observed total flux density, $B_{\nu}$ is the Planck function for a dust temperature $T_{d}$, d is the distance to the source, R is the dust-to-gas mass ratio and $k_{\nu}$ is the dust opacity which depends on the frequency as  $\kappa_{\nu}=\kappa_{o} \cdot (\nu / \nu_{o})^{\beta}$. We assume that the dust emission is isothermal and optically thin and neglect any possible effects produced by scattering. Therefore, all the masses presented should be considered as lower limits. We adopt the same assumptions as in \cite{2021_Fernandez-Lopez}, who studied the massive star-forming region G\,5.89-0.39, also associated with an explosive dispersal outflow. Hence, assuming that I16076 is a region similar to G5.89, we consider $\beta $= 2.0, a dust with thin ice mantles with 10$^{6}$ cm$^{-3}$ density, and a gas-to-dust ratio of 100. Interpolating the dust opacity models of \cite{1994_Ossenkopf} to the observed wavelenght we obtain an opacity of 1.80\,cm$^{2}$\,g$^{-1}$ at 0.9\,mm. Since we do not have a measurement of the dust temperature in the dense cores of I16076 we fix their mean dust temperature to $T_{d}=30$\,K (measured temperature for the dense clump of IRAS16076 reported by \cite{2018_Urquhart}). The estimated masses of the dense cores range from 0.3 to 22\msun (see Table \ref{tab:continuum}). To obtain the uncertainty in the masses we only considered errors in the total observed flux density. MM7, MM8 and MM9 cores are among the most massive, with masses greater than 10\msun. The masses we obtained are lower by a factor of 3 than the masses of the dense cores estimated by \cite{2021_Baug}. The reason for the discrepancies resides in the opacity, that they estimated using an expression of \cite{1983_Hildebrand}, which resulted a factor of 3 higher than ours. In addition, they warned that their larger masses had to be taken with caution, as the low angular resolution could hinder multiple systems to be marginally resolved. Such as the case of I16076-O1 in which we detect three possible dense cores (MM7, MM8 and MM9) or as I16076-O3 composed of MM1 and MM3.

\begin{deluxetable*} {cccccccc}
\tabletypesize{\footnotesize}
\tablenum{2}
\tablecaption{Submillimeter Continuum Sources in I16076 \label{tab:continuum}}
\tablewidth{0pt}
\tablehead{
\colhead{Source} & \colhead{RA} & \colhead{DEC} &\colhead{Deconvolved Size} & \colhead{Peak Intensity} &\colhead{Flux Density} & \colhead{Mass} & \colhead{Name Baug} \\
\nocolhead{} & \colhead{[ICRS]} & \colhead{[ICRS]} & \colhead{[$\arcsec\times\arcsec$, $\degr$]} & \colhead{[mJy~beam$^{-1}$]} & \colhead{[mJy]} & \colhead{[M$_{\odot}$]} & \colhead{}
}
\startdata 
MM1 & 16:11:27.758$\pm$0.008 & -51:41:55.01$\pm$0.03  & 3.8$\pm$0.2$\times$1.10$\pm$0.06, 77$\pm$1    & 4.0$\pm$0.2  & 38$\pm$2    & 2.9$\pm$0.1   & I16076$\_$O3\\
MM2 & 16:11:27.668$\pm$0.008 & -51:41:51.71$\pm$0.07  & 1.4$\pm$0.2$\times$1.1$\pm$0.2, 50$\pm$70     & 2.5$\pm$0.3  & 9$\pm$1     & 0.7$\pm$0.1   & $--$\\
MM3 & 16:11:27.660$\pm$0.001 & -51:41:55.57$\pm$0.01  & 0.87$\pm$0.01$\times$0.70$\pm$0.05, 50$\pm$20 & 26.5$\pm$0.7 & 56$\pm$2    & 4.3$\pm$0.2   & I16076$\_$O3\\
MM4 & 16:11:27.120$\pm$0.002 & -51:41:46.66$\pm$0.05  & 1.5$\pm$0.1$\times$0.5$\pm$0.2, 176$\pm$3     & 2.2$\pm$0.1  & 5.7$\pm$0.5 & 0.44$\pm$0.04 & $--$\\
MM5 & 16:11:26.870$\pm$0.007 & -51:41:55.99$\pm$0.07  & 1.5$\pm$0.2$\times$0.9$\pm$0.2, 40$\pm$20     & 5.2$\pm$0.6  & 18$\pm$2    & 1.4$\pm$0.2   & I16076$\_$O4\\
MM6 & 16:11:26.626$\pm$0.009 & -51:41:49.85$\pm$0.08  & 1.7$\pm$0.3$\times$0.9$\pm$0.2, 50$\pm$10     & 3.2$\pm$0.4  & 12$\pm$2    & 0.9$\pm$0.1   & $--$\\
MM7 & 16:11:26.550$\pm$0.001 & -51:41:59.48$\pm$0.01  & 1.36$\pm$0.03$\times$1.11$\pm$0.03, 162$\pm$6 & 18.2$\pm$0.3 & 69$\pm$2  & 5.3$\pm$0.1  & I16076$\_$O1 \\
MM8 & 16:11:26.546$\pm$0.005 & -51:41:57.23$\pm$0.06  & 1.6$\pm$0.2$\times$1.1$\pm$0.1, 20$\pm$10     & 70.4$\pm$6.0 & 290$\pm$30  & 22$\pm$1  & I16076$\_$O1\\
MM9 & 16:11:26.5392$\pm$0.0003 & -51:41:55.820$\pm$0.006  & 2.27$\pm$0.01$\times$0.981$\pm$0.008, 178.1$\pm$0.3 & 18.5$\pm$0.1 & 101.1$\pm$0.7  & 5.7$\pm$0.3  & I16076$\_$O1\\
MM10 & 16:11:26.474$\pm$0.004 & -51:41:52.64$\pm$0.03 & 2.1$\pm$0.1$\times$1.11$\pm$0.06, 58$\pm$3    & 13.9$\pm$0.6 & 74$\pm$3    & 5.7$\pm$0.3   & $--$\\
MM11 & 16:11:26.222$\pm$0.008 & -51:41:57.27$\pm$0.03 & point source                                  & 5.6$\pm$0.4  & 7$\pm$1     & 0.54$\pm$0.08 & $--$\\
MM12 & 16:11:26.12$\pm$0.01 & -51:41:59.62$\pm$0.04 & 1.1$\pm$0.3$\times$0.4$\pm$0.2, 100$\pm$10    & 2.2$\pm$0.3  & 4$\pm$1     & 0.31$\pm$0.06 & $--$\\
MM13 & 16:11:25.929$\pm$0.003 & -51:41:57.12$\pm$0.02 & 1.2$\pm$0.1$\times$0.4$\pm$0.1, 50$\pm$10     & 5.7$\pm$0.6  & 2.7$\pm$0.2 & 0.44$\pm$0.01 & $--$\\
MM14 & 16:11:25.926$\pm$0.003 & -51:41:58.32$\pm$0.04 & 1.1$\pm$0.1$\times$0.45$\pm$0.1, 160$\pm$10   & 2.9$\pm$0.2  & 6$\pm$1     & 0.46$\pm$0.05 & $--$
\enddata
\tablecomments{Masses estimated using dust temperatures of 30\,K. The last column shows the labels of the dense cores identified by \cite{2021_Baug}.}
\end{deluxetable*}

\newpage

\subsection{Outflow Identification and Kinematics} \label{sec:outflowid}

\begin{figure*}
\centering
\includegraphics[scale=0.5]{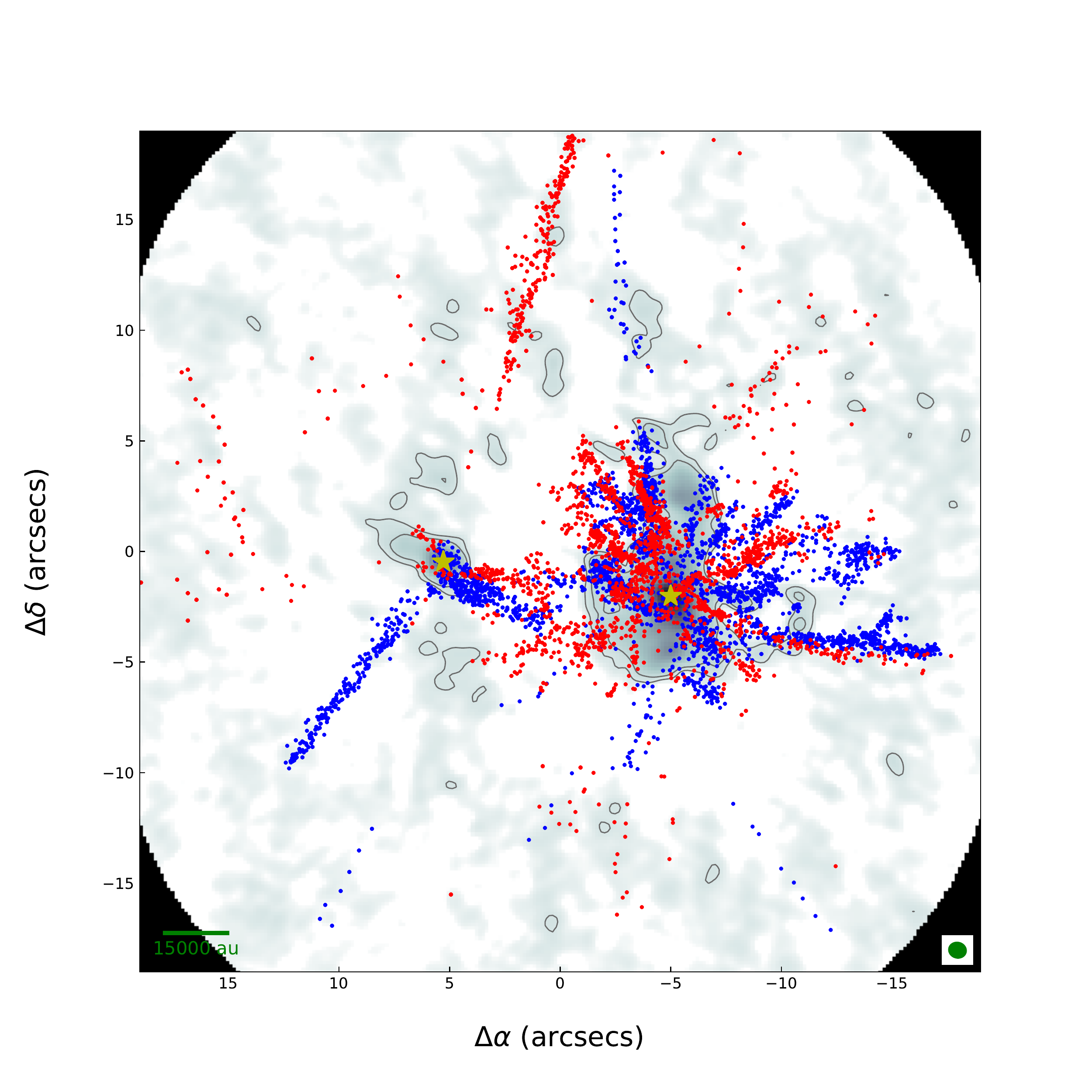}
\caption{Image of the molecular outflows in IRAS\,16076-5134. Background grey-scale image is the ALMA 0.9\,mm continuum image from Figure \ref{fig:cores} and the black contours are set at 3 and 6 times the rms noise level. The red and blue circles correspond to redshifted and blueshifted CO(3-2) gas condensation, identified in the velocity cubes as part of the outflows in the region (see Section \ref{sec:outflowid}). The driving sources of the ensemble of filaments (West) and the blueshifted monopolar outflow lobe (East) are marked with yellow stars. }
\label{fig:flujos}
\end{figure*}

\begin{figure}
\centering
\includegraphics[scale=0.55]{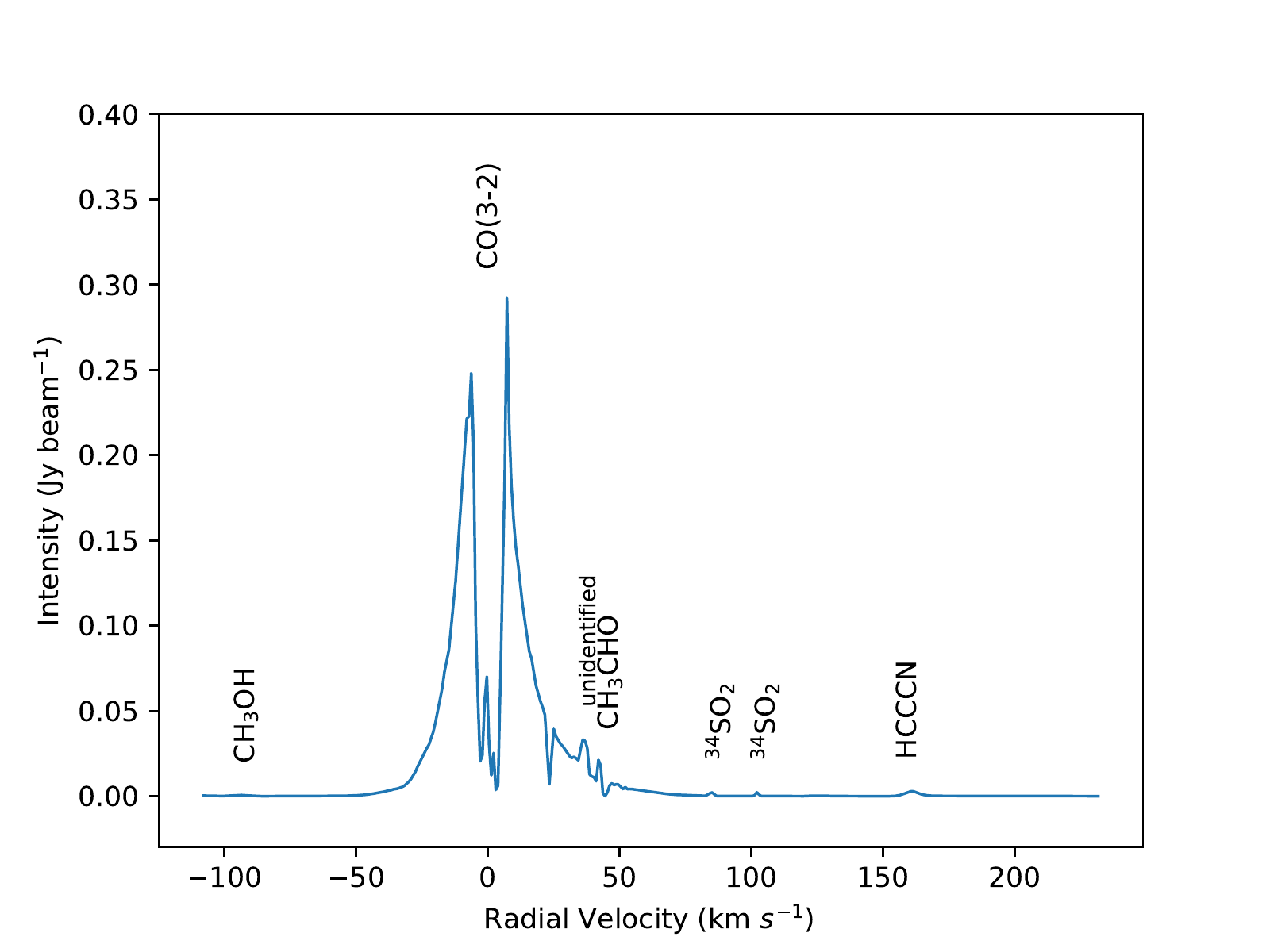}
\caption{\co spectrum integrated over the whole ensemble of filaments associated with MM8. The radial velocity is expressed relative to the system cloud velocity of $-$87.7\kms. The spectrum shows a deep absorption by the parental cloud, but also two less prominent absorptions at 23.4 and 44.5\kms, probably due to foreground clouds along the line-of-sight. There is also other spectral lines (HCCCN, $^{34}$SO$_{2}$, CH$_{3}$CHO and CH$_{3}$OH) much weaker than the main CO line.} 
\label{fig:espectro}
\end{figure}

\begin{figure}
    \centering
    \includegraphics[scale=0.5]{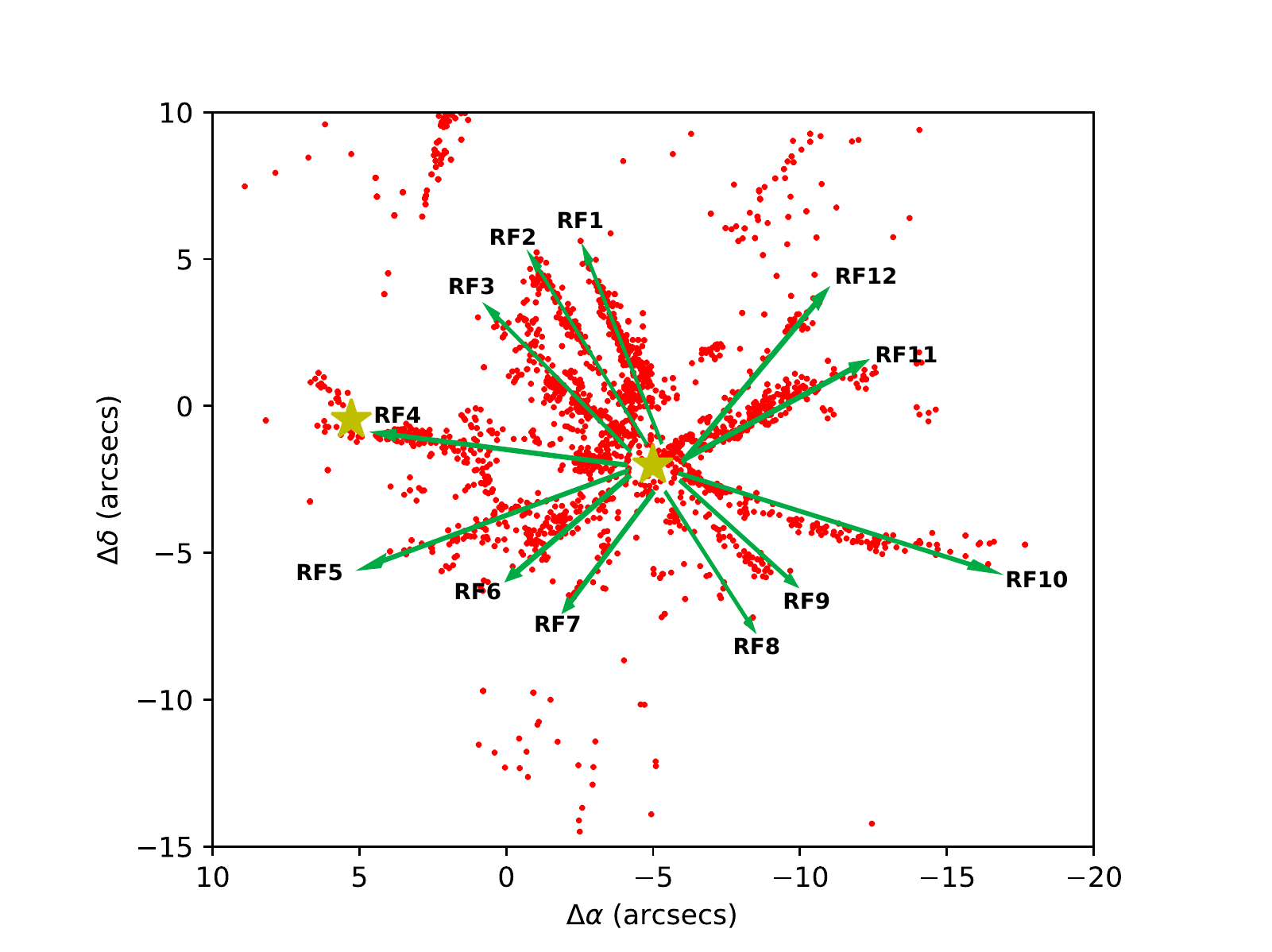}
    \hspace{1mm}
    \includegraphics[scale=0.5]{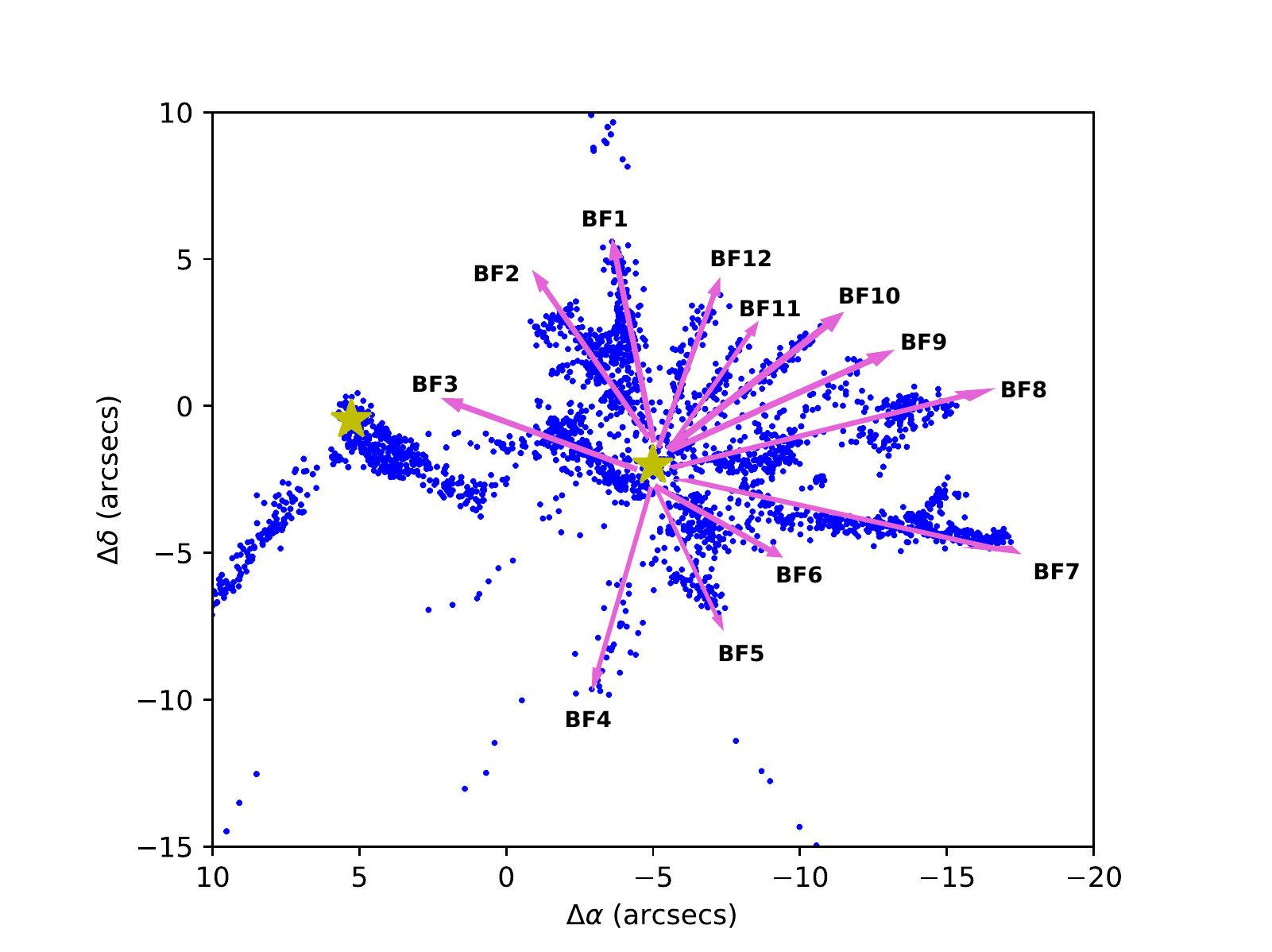}
    \caption{Redshifted (upper panel) and blueshifted (bottom panel) CO(3-2) gas condensations detected in IRAS16076. We could identified 12 redshifted filaments and 12 blueshifted, are labeled as RF and BF respectively. The posible origin of the all filamentes is marked with yellow star.} 
    \label{fig:blue-red}
\end{figure}

In Figure \ref{fig:flujos} we present the \co molecular emission in the range of radial velocities from $-$62.15 to 83.47\kms, avoiding the central velocity channels ([-5.5:6.2]\kms) that show extended CO emission from the cloud (velocities expressed with respect to the cloud velocity, v$_{sys}$=$-$87.70\kms \footnote{Throughout this paper, we work with radial (or line-of-sight) velocities, v, with respect to the cloud velocity, unless otherwise noted.}). We perform the identification of the outflow filaments using Miriad's \texttt{cgcurs}\footnote{Cgcurs displays an image displays an image and allows to interactively read image values, evaluate statistics within a poligonal region that can be interactively defined and recorded into a text file.} task \citep{Sault_1995}. We manually obtained the positions and line-of-sight velocities of all half-beam (0.5 arcsec) condensations where the CO emission, in all pixels belonging to the condensation, is greater than 3$\sigma$ (34\mjyb). Each condensation is marked with a red/blue dot in case its velocity is redshifted/blueshifted. This procedure was repeated for each channel of the CO data cube, except for the central channels where the cloud emission was predominant and the channels with absorption features at 23.4 and 44.5\kms (see Fig. \ref{fig:espectro}).  These absorption features may be due to colder and overlapping foreground molecular clouds along the line of sight, and are out of the scope of this contribution. In addition, the ALMA spectrum reveals other weak emission lines from other chemical species (CH$_3$OH, $^{34}$SO$_2$, HC$_3$N, CH$_3$CHO), which are not focus of the present work. After collecting the positions for the \co condensations we continued to building up a map in which dozens of filament-like structures can be distinguished emerging quasi-radially from a common center near the MM8 peak. Two more north-south extended monopolar filaments (one blueshifted to the south and one redshifted to the north) can be seen East of the filaments associated to MM8. The blueshited filament, with a length of $\sim11\arcsec$, seems to be associated with the MM3 core. The association of the northern redshifted filament ($\sim18\arcsec$ in length) with a dense core is not clear, although \cite{2020_Baug} attributed it to the I16076-O2 core. In this work, we focus on the analysis of the 24 radial filaments identified around MM8. We found 12 blueshifted and 12 redshifted filaments (see Figure \ref{fig:blue-red}). Most of them coincide with outflows reported by \cite{2020_Baug}, and associated with the dense core IRAS16076-O1 (resolved into MM7, MM8, MM9 in our work). Filament parameters, such as terminal velocity, coordinates of its outer tip, length and position angle (PA) are given in Table \ref{tab:parameters}. Some of the filaments show very high radial velocities (v$>$50\kms), while others are more quiescent (v$<$10\kms). They show a wide range of PA values, indicating that, if they are somehow related, they do not have a preferred direction.

%\begin{figure}
%    \centering
%    \subfigure{\includegraphics[scale=0.5]{red.eps}}
%    \hspace{1mm}
%    \subfigure{\includegraphics[scale=0.5]{blue.eps}}
%    \caption{Redshifted (upper panel) and blueshifted (bottom panel) CO(3-2) gas condensations detected in IRAS16076. We could identified 12 redshifted filaments and 12 blueshifted, are labeled as RF and BF respectively. The posible origin of the all filamentes is marked with yellow star.} 
%    \label{fig:blue-red}
%\end{figure}

\begin{deluxetable} {cccccccccccc}
\tabletypesize{\footnotesize}
\tablenum{3}
\tablecaption{Outflow parameters \label{tab:parameters}}
\tablewidth{0pt}
\tablehead{
\colhead{Outflow} & \multicolumn2c{End Coordinates} & \colhead{V$_{end}$} &\colhead{Extent} & \colhead{PA} \\
\colhead{Name} & \colhead{[RA (J2000)]} & \colhead{[Dec (J2000)]} & \colhead{[km~s$^{-1}$]} & \colhead{[arcsec]} & \colhead{[$\degr$]}
}
\startdata 
RF1 & 16 11 26.784 & -51 41 50.28 & 51.36 & 7.6 & 18 \\
RF2 & 16 11 26.976 & -51 41 50.28 & 47.98 & 7.9 & 31 \\
RF3 & 16 11 27.096 & -51 41 52.80 & 49.67 & 7.1 & 50 \\
RF4 & 16 11 27.768 & -51 41 56.04 & 11.55 & 11.8 & 85 \\
RF5 & 16 11 27.456 & -51 42 00.00 & 9.86 & 9.3 & 109 \\
RF6 & 16 11 27.168 & -51 42 01.08 & 9.86 & 7.0 & 124 \\
RF7 & 16 11 26.856 & -51 42 01.44 & 54.75 & 5.2 & 146  \\
RF8 & 16 11 26.208 & -51 42 02.52 & 6.47 & 6.2 & -150  \\
RF9 & 16 11 26.352 & -51 42 00.36 & 28.49 & 5.4 & -131  \\
RF10 & 16 11 25.320 & -51 42 00.36 & 20.02 & 12.2 & -106  \\
RF11 & 16 11 25.968 & -51 41 54.24 & 29.34 & 6.5 & -62  \\
RF12 & 16 11 25.992 & -51 41 51.72 & 9.01 & 7.9 & -43  \\
BF1 & 16 11 26.736 & -51 41 49.56 & -19.78 & 8.2 & 13   \\
BF2 & 16 11 26.880 & -51 41 51.72 & -33.34 & 6.7 & 30  \\
BF3 & 16 11 26.928 & -51 41 55.68 & -44.35 & 4.3 & 68  \\
BF4 & 16 11 26.760 & -51 42 04.32 & -14.70 & 7.7 & 164  \\
BF5 & 16 11 26.304 & -51 42 02.16 & -18.09 & 5.7 & -156 \\
BF6 & 16 11 26.256 & -51 41 59.64 & -25.71 & 4.0 & -133  \\
BF7 & 16 11 25.272 & -51 41 59.64 & -46.04 & 12.5 & -102  \\
BF8 & 16 11 25.488 & -51 41 54.96 & -33.35 & 10.5 & -77  \\
BF9 & 16 11 25.800 & -51 41 53.52 & -12.16 & 8.1 & -62  \\
BF10 & 16 11 25.968 & -51 41 52.80 & -28.25 & 7.2 & -51  \\
BF11 & 16 11 26.232 & -51 41 53.16 & -32.49 & 5.3 & -36  \\
BF12 & 16 11 26.376 & -51 41 51.72 & -20.63 & 6.0 & -16 
\enddata
%\tablecomments{caption}
\end{deluxetable}

\begin{figure*}
\centering
\includegraphics[scale=0.45]{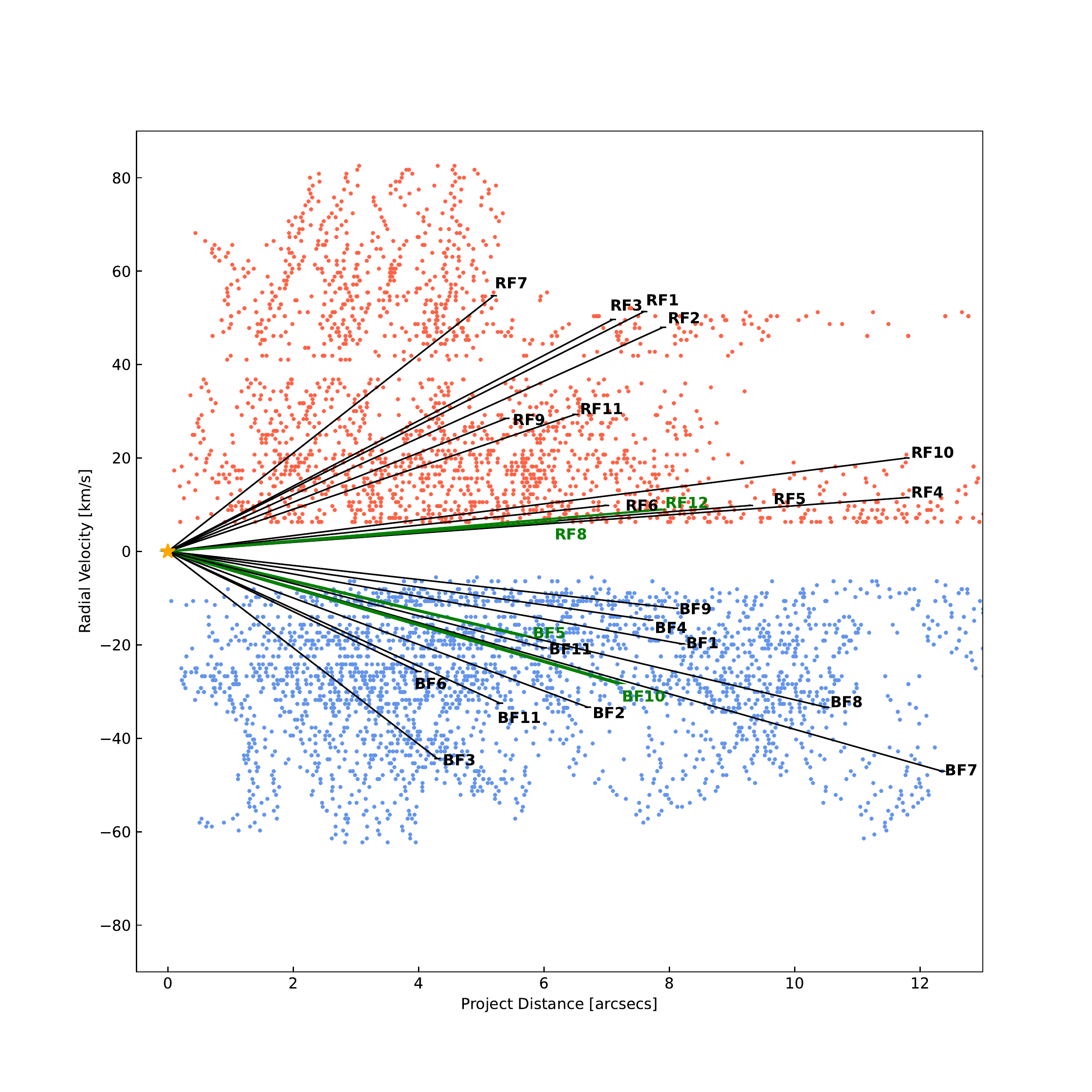}
\caption{Position-Velocity diagram of the CO(2-1) emission from the ensemble of filaments associated with IRAS\,16076-5134. The blueshifted and redshifted dots are extracted analyzing the CO condensations of the filamentary outflows emerging from the MM8 dense core. The black straight lines were plot linking the yellow star, placed at the location of MM8 at 0\kms (radial velocities given with respect to the cloud velocity) and the tip of every one of the 24 filaments of the ensemble. Better angular resolution is needed for better analysis of filament kinematics.} 
\label{fig:pv}
\end{figure*}

Figure \ref{fig:pv} shows a diagram with the line-of-sight velocity as function of the projected distance (PV diagram) of the CO molecular condensations traced by the filaments around MM8 (up to a projected distance $<$15\arcsec). The redshifted and blueshifted filaments are distinguished by red and blue dots, respectively. As said before, no molecular condensations were marked around 0\kms due to extended emission contamination. In addition, extended emission is also observed at 42.4\kms throughout the field of view. These can be explained by the presence of other emitting cloud along the line-of-sight, whose extended emission is filtered by the interferometer, hindering the spatial distribution of the filament's from I16076. The emission from filaments with gas at higher velocities appears more compact (filaments with projected lengths $<$2\arcsec) and close to the origin of the filaments, near the position of MM8. The black and green arrows join the central position (the peak of the MM8 core at 0\kms) and the tip of each filament (see Table \ref{tab:parameters}). The current angular resolution of the observations does not allow the kinematics of the filaments to be adequately analyzed from the PV diagram.

\begin{figure*}
\centering
\includegraphics[scale=1.1]{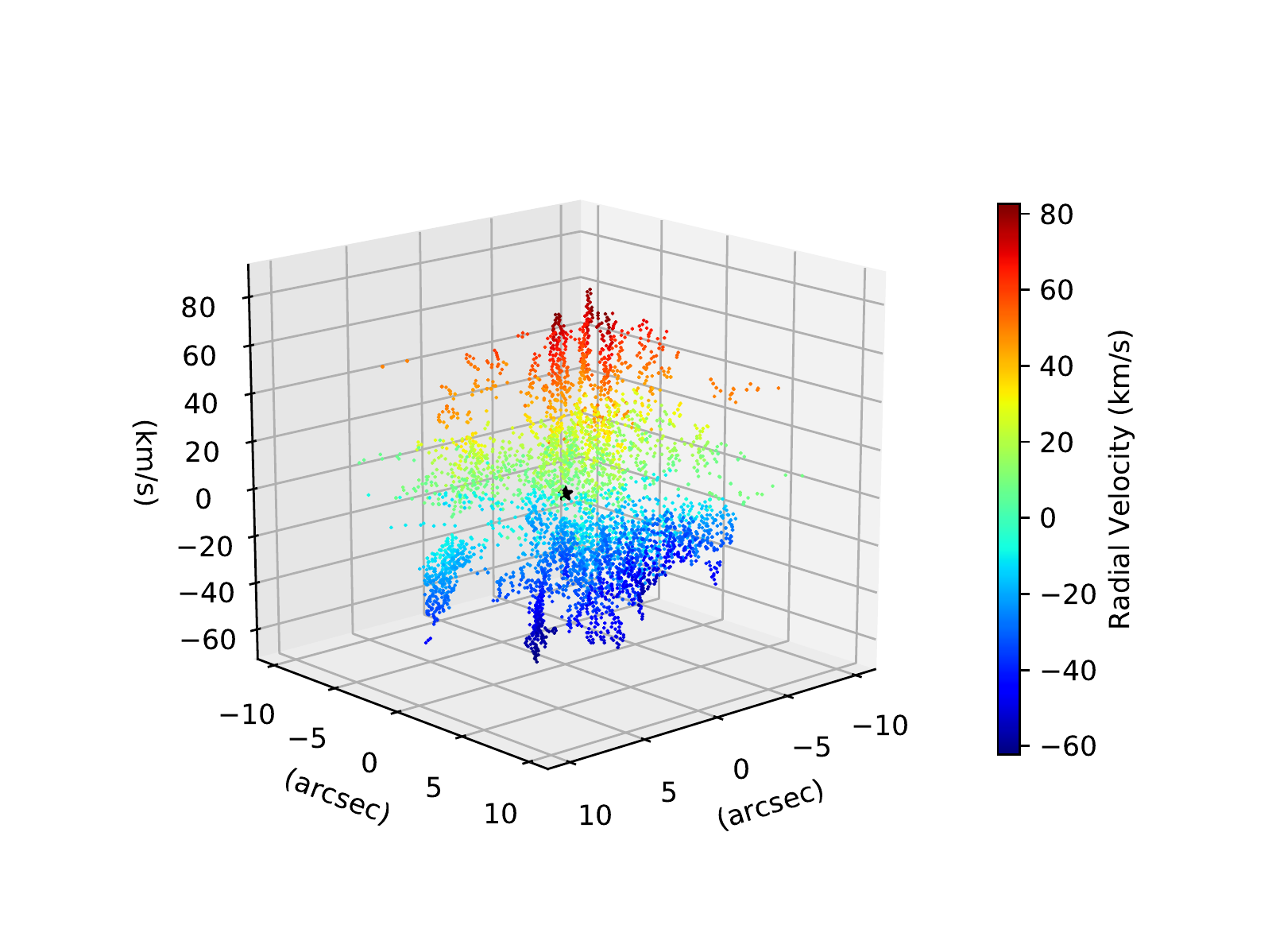}
\caption{Three-dimensional representation of the dispersal outflowing event in IRAS\,16076-5134. The radial blueshifted and redshited velocities are show from blue to red colors. The LSR radial velocity scale-bar (in km s$^{-1}$ ) is shown on the right. The ($0\arcsec$, $0\arcsec$, 0 km s$^{-1}$) position is the origin of the plot and is marked by a black star, representing the origin of the dispersal outflow.}
\label{fig:3d}
\end{figure*}

A 3D representation of the 24 \co filaments ensemble toward I16076 gives a better idea of the isotropic morphology and its ordered kinematic pattern (Figure \ref{fig:3d}). The X and Y axis in this graphic are the projected distance in RA and DEC (in \arcsec), while the Z axis represents the radial velocity (in \kms) of the filaments, which is also displayed by a color scale. The central position is at $(\alpha,\delta)_{J2000.0}=16^{h}11^{m}26.546^{s},-51^{\circ}41^{'}57.23^{''}$ (the MM8 core position). For this image to represent the actual 3D distribution of the filaments, the filaments should be undergoing an expansive motion at constant velocity following ballistic trajectories. Assuming this, the Figure \ref{fig:3d} would provide a general view of the spatial distribution and kinematics of the filaments ensemble. As can be seen in Figure \ref{fig:flujos}, \ref{fig:blue-red} and \ref{fig:3d}, one of the main characteristics of the filaments is their quasi-isotropic distribution on the plane of the sky projected. They also appear to have a common origin in space and velocity. Furthermore, the velocity within most of the filaments become bluer or redder with the distance from the origin. This indicates that the filaments presents a velocity gradient. Although a velocity gradient similar to a Hubble's law is observed, this does not imply that filaments are currently undergoing acceleration. Since no clear accelerating source is currently increasing the velocity of the filaments, these velocity gradients are likely related with the decay of the velocity of the molecular material or produced by the interaction between the filaments and the environment material (see Section \ref{sec:outflowenergy}).

\subsection{Outflow Energetics} \label{sec:outflowenergy}

In light of the findings explained in the previous Section \ref{sec:outflowid} we assume hereon that the 24 radial filament ensemble constitutes a single outflow entity with a common origin. In this Section we estimate the physical parameters of the ensemble such as mass, momentum and kinetic energy. We follow an analogous procedure to that indicated in \cite{2020_Fernandez-Lopez} in G5.89-0.39, and consider local thermodynamic equilibrium and optically thin CO emission. To derive the column density of the \co transition we use the equation:

$$N_{tot}(CO)=\frac{3h}{8 \pi^{3} \mu^{2}_{B} J} \frac{Q_{rot}\, e^{E_{u}/kT_{ex}}}{e^{h\nu/kT_{ex}}-1} \frac{\int T_B dv}{[J_{\nu}(T_{ex})-J_{\nu}(T_{bg})]} = $$

\begin{equation}
=\frac{1.195 \times 10^{14} (T_{ex}+0.922) e^{33.192/T_{ex}}}{e^{16.596/T_{ex}}-1} \frac{T_{B} \Delta v}{J_{\nu}(T_{ex})-J_{\nu}(T_{bg})}  
\end{equation}

%$$N_{tot}(CO)=\frac{3h}{8 \pi^{3} \mu^{2}_{B} J} \cdot 
%\frac{Q_{rot}\, e^{E_{u}/kT_{ex}}}{e^{h\nu/kT_{ex}}-1} \cdot
%\frac{\int T_B dv}{[J_{\nu}(T_{ex})-J_{\nu}(T_{bg})]} 
%=\frac{1.195 \times 10^{14} (T_{ex}+0.922) e^{33.192/T_{ex}}}{e^{16.596/T_{ex}}-1} \cdot
%\frac{T_{B} \Delta v}{J_{\nu}(T_{ex})-J_{\nu}(T_{bg})}\quad\cdot $$

In the expression above we use the Planck's constant $h=6.6261\times10^{-27}\,erg\,s$, the dipole moment $\mu_{B}=1.1011\times10^{-19}$\,StatC\,cm, the partition function $Q_{rot}=kT_{ex}/(hB_{0})+1/3$ with a rigid rotor rotation constant $B_{0}=57.635968$\,GHz, the quantum number of the upper level $J=3$, the Rayleight-Jeans equivalent temperature $J_{\nu}(T)=(h\nu/k)/(e^{h\nu/kT}-1)$,the brightness temperature $T_{B}$ in K and the velocity interval $\Delta v$ in\kms. Regarding the excitation temperature, we take two different values: 40 and 70\,K (Felipe Navarete, private communication; see also \cite{2019Navarete}), appropriate for many massive clumps in the Milky Way. We estimate the mass in each velocity channel as: 
\begin{equation} \label{M_CO}
M_{out}=\mu mh\Omega N_{tot}/X_{CO}
\end{equation}
%$$M_{out}=\mu mh\Omega N_{tot}/X_{CO}$$
where $\mu$ is the mean molecular weight, which is assumed to be equal 2.76 \citep{1999_Yamaguchi}, \textit{m} is the hydrogen atom mass ($\sim 1.67\times10^{-24}g$), $\Omega$ is the solid angle, and a CO abundance of X$_{CO}$=10$^{-4}$. We derive a total mass for the dispersal explosive outflow as the sum of the masses of the all filaments in each channel (see Table \ref{tab:parameters}). In order not to overestimate the mass measurement, we exclude the channels contaminated by the emission of the parental molecular cloud and the other clouds along the line-of-sight, as well as the channels affected by other molecular lines emission (CH$_{3}$CHO and unindentified) toward the wing of CO line. Finally, we add all the channel masses to derive a total mass between 138 and 216\msun, for 70\,K and 140\,K respectively.

\begin{figure}
\centering
\includegraphics[scale=0.35]{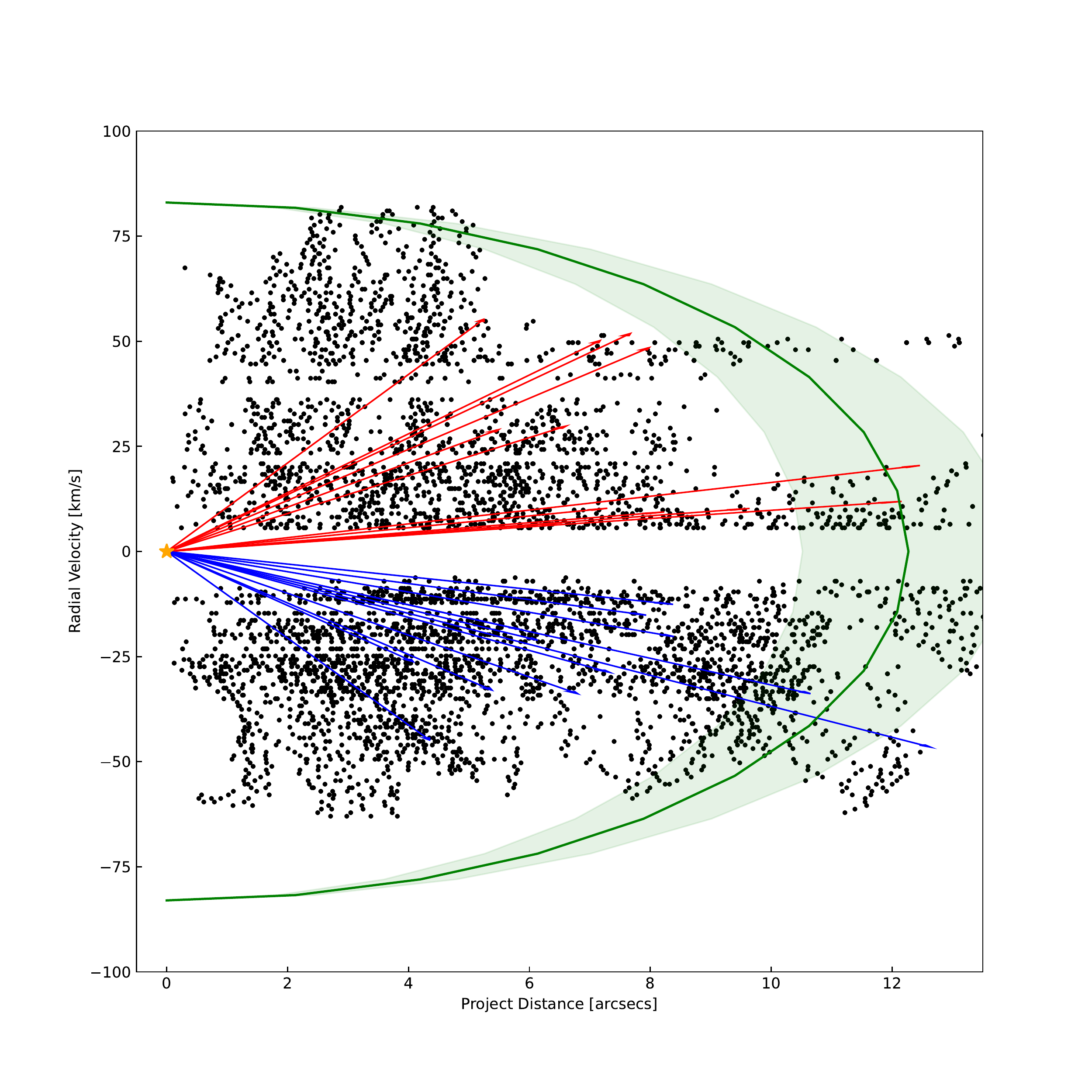}
\caption{Models (green curve and shaded region) of good qualitative agreement of an isotropic expansion at constant velocity (V$_{outflow}$=83\kms) on the PV diagram setting t$_{dyn}=3500\pm500$\,years.} 
\label{fig:ajuste}
\end{figure}

Recently, \cite{Raga_2020} presented an analytical `head/tail model' that could explain the kinematics found in the explosive Orion BN/KL flow. This model considers an ejected plasmon, where the tail of the plasmon is decelerating due to the interaction with the medium it passes through. The PV diagrams resulting from this model are similar to those of the Orion BN/KL filaments \citep{2009_Zapata,2017_Bally}, which show a velocity gradient (the radial velocity increases with projected distance). As a first approximation, we consider a simple scenario, in which the plasmons travel with no friction. Therefore, the plasmons follow constant velocity trajectories. We also assume that plasmons are ejected isotropically from a common origin. Therefore, the plasmon will follow a roughly constant velocity trajectory, and we also assume that they are ejected isotropically from the origin of the explosion. To better estimate the momentum and kinetic energy it is necessary to take into account the inclination of each filament ($v_{filament} = v_{rad}/cos(i)$, where $v_{rad}$ is the velocity along the line-of-sight and i the inclination with respect to the plane of the sky).
But the resolution of the observations does not allow us to obtain information about the inclinations of each filament, so we will take the maximum observed radial velocity (83\kms) as the expansion velocity (V$_{outflow}$) of all filaments. With this V$_{outflow}$, we can estimate a lower limit for the dynamic age, momentum and kinetic energy of the outflow. We perform this analysis in Figure \ref{fig:ajuste}, which shows a PV diagram similar to Figure \ref{fig:pv}. On the PV diagram we have superimposed models (green curve and shaded region) of an isotropic expansion at constant velocity varying its dynamic age. Models of good qualitative agreement are found manually by using $t_{dyn}=3500\pm500$\,years. Finally, the derived total outflow momentum are 1.2$\times10^4$\msun\kms and 1.8$\times10^4$\msun\kms, and total outflow kinetic energy are 9.9$\times10^{48}$\,erg and 1.5$\times10^{49}$\,erg for 70\,K and 140\,K, respectively. This implies that the outflow emission in I16076 is associated with a very energetic event.

\section{DISCUSSION} \label{sec:discussion}
\subsection{I16076 a New Explosive Dispersal Outflow?} \label{sec:explosion}
The outflow found in the high-mass star-forming region I16076 consists of an ensemble of at least 24 radial high-velocity filament-like ejections (ranging  from $-$62 to +83\kms) distributed quasi-isotropically. The filaments point back to approximately the same central position, where the MM8 dense core is located (Fig.\ref{fig:flujos}). Spatial overlap of the redshifted and blueshifted filaments is observed. The filaments presents a linear velocity gradient (see Fig. \ref{fig:3d}). In addition, the estimated energy of the filament ensemble is at least one to two order of magnitude larger than that of typical outflows associated with high-mass protostars. All these features point to a direct analogy with those of the explosive dispersal outflows: Orion-KL, DR21 and G5.89-0.39 \citep[e.g.,][]{2017_Zapata, 2016_Bally, 2020_Zapata}. The collected evidence strongly suggests that I16076 is a new explosive dispersal outflow candidate. Therefore, the ejection of filaments would have occurred from a single brief energetic event. However, new high-angular and more sensitive observations are needed to confirm it.

\subsection{Rate of Explosive Dispersal Outflows in the Milky Way}\label{sec:rate}

\begin{figure*}
\centering
\includegraphics[scale=0.6]{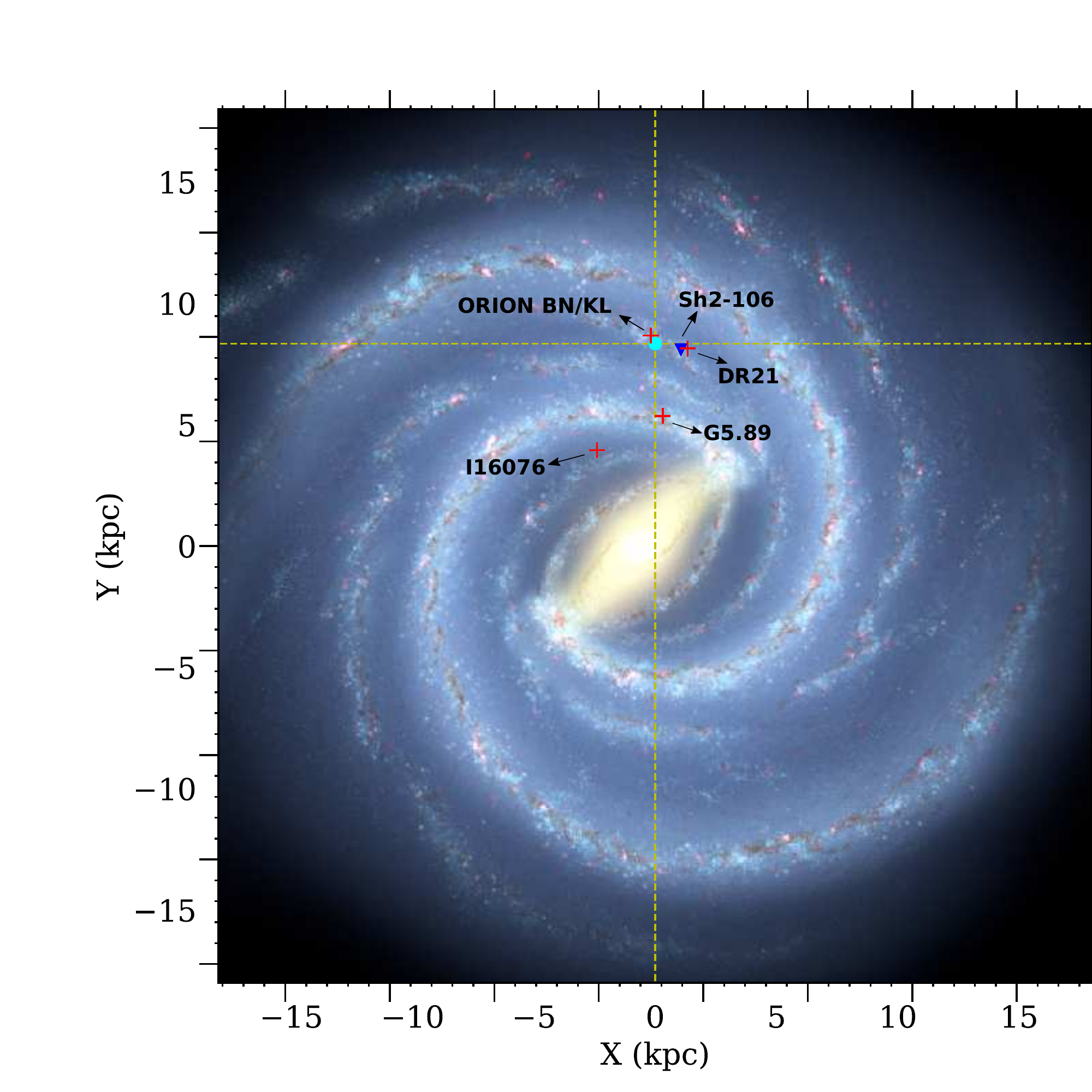}
\caption{Sketch of the projected spatial distribution of the confirmed explosive dispersal outflows (red crosses), plus S106\,IR \citep[blue inverted triangle,][]{2022_Bally} in The Milky Way (artist’s concept, R. Hurt: NASA/JPLCaltech/ SSC). The location of the Sun is marked with a circle at the crossing of the two dashed lines.}
\label{fig:MW}
\end{figure*}

As indicated by the elements explained in Section \ref{sec:explosion} the ensemble of filaments in I16076 could be a new explosive dispersal outflow. The discovery of another of such dispersal events could have further implications, that we explore in this Section. For instance, we do not know accurately how common these dispersal outflows in the Galaxy, if they play a role in the energy budget of the Interstellar Medium, or the physical processes that originate them. Figure \ref{fig:MW} shows the spatial distribution of the currently confirmed dispersal outflows in the Milky Way, along with I16076 (red crosses). S106\,IR\footnote{This region has a complex of H$_2$ and N[III] compact structures with increasing proper motions with projected distance from the young high-mass stellar object, but the nature of these features still needs further confirmation.}, a recently reported possible explosive dispersal outflow \citep[inverted blue triangle,][]{2022_Bally} is also included in the sample. Making a crude estimate, \cite{2020_Zapata} derived a frequency rate of one explosive dispersal event every 130\,years in the Milky Way, considering Orion\,BN/KL, DR21 and G5.89-0.39. Assuming that I16076 and S106\,IR are explosive dispersal outflows, we can refine this rate estimate. First, assume that the explosive dispersal outflows are evenly spaced in time during the last 15330\,years (it is the lapse of time between all 5 outflows, taking into account their different distances to Earth). Second, the dispersal outflows are within a projected circle of 5.6\,kpc in diameter (the separation between I16076 and DR21, which are the most separated pair of known dispersal outflows). Finally, extrapolate the frequency rate to the disk of the Galaxy, which we assume as a flat disk with a 15\,kpc radius. Since the star formation rate is fairly constant with galactocentric distance (e.g., \cite{2014_Urquhart,2006_Nakanishi}), and we only have a few explosive outflow cases, the frequency rate obtained will be a crude rate of first order approximation. Under these assumptions, the estimate of the frequency of dispersal outflows in the Galaxy should be regarded as a lower limit. In this manner, we obtain a rate of one dispersal outflow every 110\,years. This frequency rate could be even larger with the availability of a more complete high-resolution and high-sensitivity survey (not just targeted observations), which might lead to the identification of more explosive events within the considered radius.

It is worth to note that this frequency rate is comparable to the rate of supernovae \citep[one every 50 years][]{2006_Diehl}. This fact may have implications for the energy budget of the Galaxy, even accounting for the $\sim100$ times lower amount of energy (10$^{49}$\,erg) that a dispersal outflow deposits in the Interstellar Medium compared with a core-collapse supernovae that releases energies of 10$^{51}$\,erg \citep{2003_Hamuy}. The origin of the explosive dispersal outflows is probably associated with the disruption of a massive young non-hierarchical stellar system, and/or with a protostellar collision \citep{2009_Zapata, 2011_Bally, 2013_Zapata, Rivilla_2014,2016_Bally,2017_Bally}. One of its consequences is the ejection of a small group of runaway protostars (e.g., \cite{Gomez_2008,Rodriguez_2020}). Hence, there seems to be a link between the origin of the explosive dispersal outflows and the stellar dynamical interactions. The rate of dispersal explosive outflows would show that dynamic interactions and stellar collisions may be a common scenario in high-density clusters that eject runaway stars. Hence, a more complete survey of these kind of outflows may be important to advance our understanding of how massive stars form.
\newpage

\section{CONCLUSIONS}  \label{sec:conclusion}
In this paper we presented archival observations of the 0.9\,mm continuum and \co spectral line emission toward the massive star-forming region IRAS\,16076-5134 carried out with ALMA. We found 14 dense cores associated in two small clumps. The cores have masses in the range from 0.3 to 22\msun. We found an ensemble of 24 radial filament-like ejections of \co line emission (12 blueshifted and 12 redshifted) ranging from $-$62.15 to +83.47 \kms respect to the cloud velocity (-87.7\kms). This filaments are apparently emerging from the neighborhood of the dense core MM8. Also, we reported two more extended North-South monopolar outflows. One blueshifted associated with MM3, and one redshifted launched from an unidentified source. 

We analyzed the \co morphology and kinematics of the filament ensemble around MM8. The size of the ensemble of filaments is $\sim$20\arcsec and each filaments are distributed in a radial fashion, quasi-isotropically around a common center located close to the peak emission of MM8. The velocity of filaments shows a velocity gradient. The velocity increases further away from the common center, as it happens with the fragments of an explosion. Using the integrated \co emission we calculate a total mass for the filaments between 138-216\msun. Assuming an isotropic expansion model with constant velocity of 83\kms and dynamical age of 3500 years, we derive a rough estimate for the momentum and kinetic energy of the gas of the ensemble of filaments. The order of magnitude of the kinematic energy is $10^{48-49}$\,erg. Based on the evidence, we suggest that the ensemble of filaments in IRAS\,16076-5134 is an explosive dispersal outflow candidate. Under this hypotheses, we made an raw estimate of the lower limit of the rate of dispersal outflows in the Galaxy considering constant star formation rate and efficiency with respect to the galactocentric radio of the Galaxy. The value obtained is comparable to the rate of supernova explosions, which may have further implications on the role of the energy balance of the dispersal outflows in the Interstellar Medium, and the nature and origin of this kind of outflows in massive star formation. 
\\

\newpage

\section*{Acknowledgements}
This paper makes use of the following ALMA data: ADS/JAO.ALMA\#2017.1.00545.S. ALMA is a partnership of ESO (representing its member states), NSF (USA) and NINS (Japan), together with NRC (Canada) and NSC and ASIAA (Taiwan), in cooperation with the Republic of Chile. The Joint ALMA Observatory is operated by ESO, AUI/NRAO and NAOJ. We thank the anonymous referee for the valuable comments and suggestions that help to improve the manuscript. EGC acknowledges financial support from CONACyT-
280775 and thanks the hospitality of the IRyA (CONACIT), Morelia, Mexico. TB acknowledges the support from S. N. Bose National Centre for Basic Sciences under the Department of Science and Technology, Govt. of India.

%\newpage

\bibliography{I16076}{}
\bibliographystyle{aasjournal}

\end{document}